# What are the key components of an entrepreneurial ecosystem in a developing economy? A longitudinal empirical study on technology business incubators in China

Xiangfei Yuan[a,c], Haijing Hao[b,c], Chenghua Guan[d], Alex Pentland[c]

[a] Capital Institute of Science and Technology Development Strategy, Beijing, China

[b] Computer Information Systems Department, Bentley University, Waltham, MA, USA

[c] Media Lab, MIT, Boston, MA, USA

[d] Beijing Normal University, Beijing, China

**Abstract**

Since the 1980s, technology business incubators (TBIs), which focus on accelerating businesses through resource sharing, knowledge agglomeration, and technology innovation, have become a booming industry. As such, research on TBIs has gained international attention, most notably in the United States, Europe, Japan, and China. The present study proposes an entrepreneurial ecosystem framework with four key components, i.e., people, technology, capital, and infrastructure, to investigate which factors have an impact on the performance of TBIs. We also empirically examine this framework based on unique, three-year panel survey data from 857 national TBIs across China. We implemented factor analysis and panel regression models on dozens of variables from 857 national TBIs between 2015 and 2017 in all major cities in China and found that a number of factors associated with people, technology, capital, and infrastructure components have various statistically significant impacts on the performance of TBIs at either national model or regional models.

**Keywords**

entrepreneurial ecosystems, technology business incubator, innovation, panel regression, factor analysis, national model, regional model

## 1.    Introduction

Entrepreneurial ecosystems have been widely studied in recent years as one of the most appealing topics for industrial practitioners, government policymakers, and academic scholars (Spigel, 2017; Roundy et al., 2017; Autio et al., 2018). Many studies have investigated the nature, networks, institutions, and dynamics of entrepreneurial ecosystems to improve the understanding on how to support the entrepreneurship growth (Chen et al., 2019; Roundy et al., 2018; Mack & Mayer, 2016). Stam and Spigel (2016) discussed an integrative model that consists of functional attributes, social and physical conditions of human interaction, and crucial elements such as networks of entrepreneurs, leadership, talent, knowledge, and support services of entrepreneurial ecosystems. However, the authors also pointed out that most entrepreneurial ecosystem studies were qualitative case studies, lacking in quantitative research.

Other research also revealed a dearth of empirical evidence for evaluating the performance of entrepreneurial ecosystems because of the current reality, i.e., no data for the necessary variables (Spigel, 2017; Chen et al., 2019). Another reason for lacking this stream of empirical quantitative research is the complexity and vague boundaries of entrepreneurial ecosystems, which are difficult to isolate from regional innovation systems; further, the regional social-economic context has enhanced the difficulty and complexity of quantitative studies. Spigel (2017) believed that the entrepreneurial ecosystem's boundary should be beyond a firm but within a region. In other words, in order to study an entrepreneurial ecosystem's performance, we need to determine the proper unit of analysis as an entrepreneurial ecosystem. Should it be at a country, state, or city level? Or a business incubator level (Miller, 2017; Colombo et al., 2019)? After extensive exploration studies, we believe a business incubator is the smallest and atomic entrepreneurial ecosystem because one incubator includes many actors, such as entrepreneurs, mentors, service staff, and investors, who share contextual conditions such as basic infrastructure, local economy, and government policies, which consist of an ecosystem with clear boundaries.

The concept of a business incubator started in the 1950s and has maintained worldwide popularity in entrepreneurship development (Mian et al., 2016). A business incubator is a company that assists novel startup companies to grow by providing supportive services such as office space, management training, and professional staff support. The present study is more interested in technology-focused business incubators because the fastest-growing businesses are technology-related. Six out of 10 top billionaires on the Forbes list are technology entrepreneurs (Forbes, 2020), e.g., founders of Microsoft, Facebook, Amazon, Google, etc. A technology business incubator (TBI) is an organization that aims to support and accelerate the growth and success of small and medium startups, which center on applying novel technologies to industrial applications through an array of business support services such as resource sharing, knowledge agglomeration, and technology innovation (Hansen et al., 2009). Since the 1980s, technology business incubators have become a booming industry around the world because information technology and other high technologies have been the fastest development business as well as the largest corporations. Research on TBIs has gained attention from industrial practitioners, academic researchers, and policymakers worldwide, including the United States (Mian, 1996, 1997; Rice, 2002; Harper-Anderson & Lewis, 2018; Olkiewicz et al., 2019), Europe (Bergek & Norrman, 2008; Albort-Morant & Oghazi, 2016; van Weele et al., 2017; Lukeša et al., 2018), Japan (Tsai et al., 2017; Fukugawa, 2018), and China (Xiao & North, 2017, 2018; Li & Liu, 2018; Hong et al., 2016). Although scholars may have different perspectives on the effect of incubators, incubators are indeed a kind of micro-entrepreneurial ecosystem (Isenberg 2010; Amezcua, 2013; Motoyama & Knowlton, 2016; Theodoraki & Messeghem, 2017; Theodoraki et al., 2018). TBIs are an ideal model as a miniature of an entrepreneurial ecosystem, which consists of most, if not all, of the major elements, such as entrepreneurs, supportive services, infrastructure, capital, and their networks and interactions, of an entrepreneurial ecosystem (Dagnino, 2015; Malecki, 2017).

Based on the discussion of previous research on entrepreneurial ecosystems, the present study develops a new framework with four key components to evaluate the performance of entrepreneurial ecosystems at the business incubator level and, further, empirically verifies the framework with a unique longitudinal data set from China, the largest developing economy and second-largest economy in the world. Based on the empirical results from this longitudinal data set, we examine which components may have a statistically positive impact on the performance of a technology business incubator; we also analyze and verify the reliability of the framework and its influence on the success of startups as well as propose policy implications for practitioners in the innovation domain.

The rest of this paper is organized as follows. Section 2 reviews the theoretical background and hypotheses development of our study. Section 3 describes our data and research methods. Section 4 presents the empirical results. Section 5 discusses the findings and limitations as well as future research opportunities. Section 6 is the conclusion.

## 2.    Theory and Hypotheses Development

### 2.1 Entrepreneurial Ecosystems

In recent years, entrepreneurial ecosystems have emerged as an important theoretical framework that examine and explain the development of regional entrepreneurship and the success of novel startups (Isenberg, 2010; Spigel, 2017; Chen et al., 2019; Theodoraki & Messeghem, 2017; Theodoraki et al., 2018). An ecosystem's original meaning is that of a biological community of interacting organisms and their physical environment. Moore (1996) was among the first studies to introduce the concept of ecosystems to business literature and define a business ecosystem as an economic community supported by a group of interacting organizations and individual actors within the community. Isenberg (2010) further discussed entrepreneurial ecosystems and defined "entrepreneurial ecosystems" as an organic system that included a group of tangible and intangible elements, such as customers, capital market, leadership, and culture, which are organized in complex ways to interact with venture creation and entrepreneurship development.

Further, Stam (2015) stated that an entrepreneurial ecosystem is a set of interdependent actors and coordinated factors that lead to productive entrepreneurship in a particular region. Some literature focused on the elements and their interaction activities (Davis, 2016; Adner, 2017; Spigel, 2017; Stam & Spigel 2016), while others focused on the relationships or network among the elements (Isenberg, 2010; Thompson et al., 2018; Neumeyer & Santos, 2018). Some studies developed frameworks for entrepreneurial ecosystems and tended to list all the possible related elements for the ecosystem but have little empirical evidence to verify their concepts (Chen et al., 2019). Chen et al. (2019) systematically reviewed and discussed how to gain a comprehensive understanding of an entrepreneurial ecosystem based on the last two decades' research from top international and Chinese journals; Chen et al. further suggested that studies on Chinese entrepreneurial ecosystems should develop their own indigenous approaches to ensure the model is the actual Chinese model because of China's special history, culture, social structure, and management practices.

Cavallo et al. (2018) examined the literature on entrepreneurial ecosystems and developed a set of guidelines on how to understand entrepreneurial ecosystems for scholars and practitioners; the authors also pointed out that, when policymakers and researchers confronted the huge and complex subjects and background of a regional entrepreneurial ecosystem, identifying and understanding the main subsystems of an entrepreneurial ecosystem were more promising. Also, a number of studies examined the entrepreneurial ecosystem by decomposing it into subsystems, such as a university entrepreneurial ecosystem, which had less complexity and clearer boundaries and, more importantly, which made the complex entrepreneurial ecosystem a feasible subject for empirical studies (Yearworth, 2010; Theodoraki & Messeghem, 2017; Simatupang et al., 2015; Isenberg, 2010; Miller & Acs, 2017; Huang-Saad et al., 2016).

We believe that investigating an entrepreneurship ecosystem at an incubator level, for a regional economy or a national economy, is a practical approach for several reasons. First, when a region or a country is big, such as the United States, China, or India, it means a country covers a large geographic area with a diversified culture and/or uneven levels of economic development, studying an entire country as one entrepreneurial ecosystem would be too complicated. Second, in recent years, many countries or regions have established technology business incubators to nurture and accelerate entrepreneurship development, and each incubator has its own physical space, leadership, supportive services, network, infrastructure, capital channel, and consumer population. In sum, each incubator itself is a kind of miniature entrepreneurial ecosystem within a region. Last, at the incubator level, data collection, and value measurement are more approachable and practical than at the region or country levels. Hence, the present study intends to determine the essential elements for a successful entrepreneurial ecosystem, using each incubator as a micro-entrepreneurial ecosystem.

Isenberg (2010) identified 13 essential elements of an entrepreneurial ecosystem: leaders, governments, culture, success stories, knowledge, capital, nonprofits and industry associations, educations institutions, infrastructure, geographic locations, networks, venture-oriented professionals, and potential customers. Feld (2012) pointed out nine attributes that were crucial to a successful entrepreneurial ecosystem: leadership, intermediaries, network density, government, talent, support services, engagement, companies, and capital. Spigel (2017) reviewed cluster theory, innovation systems, geography, social capital, and networks and concluded three types of attributes, culture, social, and material attributes, and materialized them into 11 important attributes for an entrepreneurial ecosystem: supportive culture, histories of entrepreneurship, work talent, investment capital, networks, mentors and role models, policy and governance, universities, support services, physical infrastructure, and open markets. Chen et al. (2019) examined 85 research articles and extracted 12 widely accepted elements, which either overlapped with or were similar to those of Isenberg (2010) and Spigel (2017). We can see that the essential elements from the prior studies by different scholars have a lot overlaps or commonalities, e.g., they all included elements of culture, government, infrastructure, and leadership, etc.

Isenberg (2010) claimed that leaders and venture-oriented professionals are related to people or human capital of an entrepreneurial ecosystem and that educational institutions, infrastructure, and geographic locations are related to or determined by economic or social infrastructure. Similarly, Spigel (2017) discussed elements, such as mentors and supportive services, related to human capital, universities and physical infrastructure, which are related to economic or social infrastructure, although the specific terms or phrases used in the studies may differ. The World Economic Forum (2013) stated that local and international markets, human capital and financing, mentorship and support systems, robust regulatory frameworks, and major universities are an ecosystem's most important pillars.

Further, we found that Isenberg (2010), Spigel (2017), and other studies missed one important attribute: technology. We believe that technology is an important element of an entrepreneurial ecosystem because, first, technology is an important driver behind the entrepreneurship development in the last several decades. Second, in recent years, many business incubators have focused on nurturing technology innovation, and they are technology business incubators and not traditional business incubators. Also, as we mentioned earlier, six out of 10 top billionaires on the Forbes list are technology entrepreneurs (Forbes, 2020), which shows how the current business is driven by technology.

Thus, we develop a set of essential elements for the entrepreneurial ecosystem for technology business incubators, categorizing them as four key components: people component, technology component, capital component, and infrastructure component.

## 2.2 Development of Entrepreneurial Ecosystems' Key Components

Based on extensive studies of theoretical and empirical research, we propose four key components for a technology-centered entrepreneurial ecosystem, and each component may have its own set of attributes to embody the ecosystem's elements and interrelationships. The four components are people, technology, capital, and infrastructure. The people component includes the human capital of an entrepreneurial ecosystem, e.g., mentorship, leadership, and supportive services, which are provided by incubators and are essential for incubatee startups to grow. The technology component is an incubator's innovation, which plays a significant role in the establishment and development of technology startups. Most technology startups are founded by scientists and engineers who own certain novel or advanced technologies and who wish to commercialize them. The capital component refers to various capitals or subsidiaries that an entrepreneurial ecosystem receives or owns, which is an essential element to drive the success of a startup firm. The infrastructure component is the general economic and social infrastructure of an ecosystem at the incubator level, which can include internal infrastructure attributes, such as office space, labs, and external attributes such as regional economy, industrial diversity, and education institutions near the incubators.

*2.2.1 People Component*

The people component is a critical human capital, which has been shown to be critical to entrepreneurship development in the extant literature. Schwartz and Hornych (2010) argued that incubator management is crucial in establishing mutual network relationships between incubated firms. Vanderstraten and Matthyssens (2012) interviewed incubator managers, tenant managers, and experts to investigate how business incubators can differentiate themselves in the incubator market. They found that certain services, in-depth operational business support, administrative services, and personal network contacts by the incubators can strategically position the incubators in the market. Van Weele et al. (2017) found that entrepreneurs may not use an incubator's resources if they believe that the incubator manager is inexperienced, which means high-quality incubator managers or service people are critical. Thus, we formulate our first hypothesis as follows:

Hypothesis 1: High-quality/quantity human capital of an entrepreneurial ecosystem will have a positive impact on the performance of an entrepreneurial ecosystem.

*2.2.2 Technology Component*

Technology innovation has been one of the most significant drivers that has inspired entrepreneurship development across the world over the last several decades (Branstetter, 2017; Branstetter et al., 2018). We want to evaluate an entrepreneurial ecosystem's innovation. Research and development (R&D) investment is a common measure used to evaluate a firm's innovation (Smith, 2006). Similarly, we can use R&D investment as an important element to measure an entrepreneurial ecosystem's innovation.

Hypothesis 2: The total R&D investment of an entrepreneurial ecosystem has a positive impact on an ecosystem's performance.

In a technology incubator, all incubatees are startup firms that focus on how to apply novel technologies, either information technology or biological technology, to production in order to grow the novel business. A patent is a form of intellectual property that gives a patent owner the exclusive right to make, use, or sell an innovative technology or a unique service to win over their competitors in the same field. Hence, another common measure used to evaluate or quantify technology innovation is the number of patents or intellectual property applications by a country or an ecosystem (Smith, 2006). We hypothesize that the number of patents or intellectual property (IP) applications is associated with an entrepreneurial ecosystem's performance. Accordingly, we formulate the next hypothesis as follows:

Hypothesis 3: The number of patents/patent applications that an entrepreneurial ecosystem owns has a positive impact on an ecosystem's performance.

*2.2.3 Capital Component*

In economics, capital refers to the assets that can improve one's power to perform economic work. Venture capital is a type of private equity that investors use to back novel startup companies that are believed to have long-term high-growth potential (Investopedia, 2020). Venture capital is the most appealing form of capital to many entrepreneurs, and many entrepreneurial success stories owe their growth to the early-support investment from venture capitalists. Thus, we believe capital should be an important component in an entrepreneurial ecosystem because, first, capital can promote startup companies' grow; second, venture capital also indicates the strong potential of a startup companies' success since most venture capitalists are experienced business men who have invested in many startups in certain domains.

Previous studies have revealed that government capital support played various roles in support of the development of incubators, e.g., financial support, tax reductions, even direct investment as stakeholders (Chandra, 2007; Pergelova & Angulo-Ruiz, 2014; Amezcua et al., 2013; Mian et al., 2016; Guan & Yuan, 2018); findings from those studies are inconsistent, however. Therefore, we seek to investigate whether various capital supports or subsidies from the government would have positive impacts on an ecosystem's performance. Thus, we formulate the hypothesis as follows:

Hypothesis 4: Capital support from venture capitalists to an entrepreneurial ecosystem has a positive impact on an ecosystem's performance.

Hypothesis 5: Financial support from governments in an entrepreneurial ecosystem has a positive impact on an ecosystem's performance.

*2.2.4 Infrastructure Component*

Infrastructure includes the basic physical and organizational structures, e.g., buildings, roads, power supplies, which are needed for the operation of a society or an enterprise; economic infrastructure supports a region's economic development, e.g., roads, rail roads, highways, water distribution networks, sewer systems, etc.; social infrastructure supports social services that can improve social comfort and connections such as schools, parks, hospitals, sports area, etc. (Torrissi, 2000). We use a region's industrial development level, such as a region's GDP, or industry diversity of a region as an indicator of economic infrastructure (Barro & Sala-i-Martin, 1995; Quigley, 1998; Chong et al., 2020) and the number of universities in the region of a technology business incubator as an indicator of social infrastructure.

Existing research has found that the local economic environment is believed to play a considerable role in the success of technical business incubators (Pe'er & Keil, 2013; Harper-Anderson & Lewis, 2018). Regional wealth, the agglomeration of knowledge-intensive business services, and the availability of venture capital are positively associated with the level of a region's economic development (Wu et al., 2007; Chi, 2008; Pan et al., 2016). Jacobs's theory on city development (Jacobs, 1996) assumed that the diversity of industries can promote innovation and regional development; this phenomenon has been demonstrated in many countries (Boschma & Iammarino, 2010; Wang, 2016; Fritsch et al., 2018; Haqa & Zhu, 2018; Content et al., 2018). Prior studies examined the relationship between industry diversity and the survival of new firms, but the results are mixed. Some researchers identified a positive association in the United States (Renski, 2011) and Italy (Basile et al., 2016), while others found no association in China (Howell et al., 2018) and Germany (Ebert et al., 2018). We develop the following hypothesis for the relationship between economic infrastructure and an ecosystem's performance.

Hypothesis 6: A better economic infrastructure of an entrepreneurial ecosystem will have a positive impact on an ecosystem's performance.

Educational institutions are regarded as part of social infrastructure (Pradhan & Rawlings, 2002). Prior research also revealed that technology startups are more likely to develop a link with educational institutions (Colombo & Delmastro, 2002; Phillimore, 1999). However, some research has shown that the interaction between incubated firms and local universities was limited and not always successful (Quintas et al., 1992; Vedovello, 1997; Bakouros et al., 2002; Westhead & Storey, 1995; Rubin et al., 2015). Thus, we developed another hypothesis to examine whether the presence of universities has an impact on the performance of incubators, as follows:

Hypothesis 7: A better social infrastructure of an entrepreneurial ecosystem will have a positive impact on an ecosystem's performance.

## 3.    Data and Methods

### 3.1 Data

The present study intends to quantitatively examine how the four key components, i.e., people, technology, capital, and infrastructure, have an impact on the performance of an entrepreneurial ecosystem, a technology business incubator in the present study, which is a mini model of an entrepreneurial ecosystem, by utilizing a large panel data set from China, which comprised of three waves of survey of more than 2,000 incubators from about 300 cities across China, spanning from 2015 to 2017, with more than 80 variables of each survey. The data were taken from the annual survey conducted by the Ministry of Science and Technology of China (MoST), 2015 to 2017, covering all state-level TBIs across China. All of the CEOs or directors of those TBIs were asked to fill the survey or double-check the survey answers if the survey was filled by their staff members to ensure the data quality and accuracy.

Over the past 30 years, technological innovation has been a critical strategy of the Chinese government in the country's economic reform and development policy, and TBIs are expected to play a primary role in this strategy (Zhang, 2017). However, the actual development of incubators did not begin until the Mass Entrepreneurship and Innovation Initiative was proposed in 2014, almost 30 years after the first TBI, the Wuhan Donghu New Technology Entrepreneurship Center, was launched in 1987 by the Chinese government to facilitate knowledge transfer from universities to industry production. Figure 1 shows the number of TBIs in China from 1995 to 2017. We can see the number increasing sharply from 2014 (1,748 TBIs) to 2017 (4,069 TBIs).

Figure 1. Number of TBIs in China from 1995–2017

TBIs play a critically influential role over entrepreneurship development in China, which had no private businesses or entrepreneurships and was a completely state-owned central planning economy before the economic reform in the early 1980s. Since the 1990s, approximately 133,000 small and medium enterprises (SMEs) have been incubated, and 24.8% of them acquired more than RMB 14.8 billion (about USD 2.19 billion) of venture capital in total (Yan, 2017). Further, 16.7% of those SMEs were listed on the Chinese Second Board (GEB) in 2016, which is a NASDAQ-style subsidiary of the Shenzhen Stock Exchange (Yan, 2017).

How to evaluate and improve the performance of business incubators has remained a longstanding question in innovation studies. Mian et al. (2016), for example, conducted a systematic literature review of technology business incubation research over the past several decades and found that only a small portion of research focused on how to evaluate the performance of business incubators. This is because assessing an incubator's performance is challenging (Hackett & Dilts, 2004; Albort-Morant & Ribeiro-Soriano, 2015). Further, Voisey et al. (2006) provided an overview of different measures of business incubator performance and suggested that sales revenue, employment growth, and the ability to graduate from an incubator were the most commonly used measures. Thus, we plan to use the number of graduated firms from an incubator as a measure of an incubator's performance, or an entrepreneurial ecosystem, which means the more venture firms can graduate from an incubator, the more mature the incubator is.

Incubators across China are heterogeneous due to widespread geographic locations and the varying levels of regional economic development. Additionally, every year, new incubators are established, and some die out with a fast-changing pace. Therefore, we want to focus on those incubators that have survived all three years of the survey period and are located in provincial capital cities or subprovincial-level cities, which have relatively broader and stronger industrial bases, better economy, and larger populations. Based on the above conditions, we narrowed down our data set to 857 state-level TBIs in 33 provincial and subprovincial cities from 28 provinces in China, with the exception of Hainan, Qinghai, and Tibet. Further, Liaoning, Shandong, Jiangsu, Guangdong, and Fujian provinces have two provincial or subprovincial cities remaining in the data set.

We further combined data from other sources beyond the data of MoST surveys, e.g., local GDP and local employment data from the China City Statistical Yearbook 2015–2017 and the number of universities in each city from China's Ministry of Education. However, local GDP is an aggregated value of the overall industry, which cannot represent the local industry's diversity, such as estimating how many sections or the size of each section. Davies and Tonts (2010) used Shannon's diversity index (Shannon, 1948) to calculate an industry's diversity index. We followed Davies and Tonts (2010) and calculated an industry diversity index H for each TBI based on the employment data of the city in which they are located, as follows:

$$H = -\sum_{i=1}^{s} p_i \ln p_i \; ,$$

where $p_i$ is the proportion of the number of people employed in industry section category $i$ and in total for $s$ categories.

Meanwhile, entrepreneurial ecosystems are contingent on regional culture and social-economic backgrounds. China is one of the largest countries by area, covers a wide range of geographic locations, contains a strong regional culture, and has various levels of economic development across regions; hence, researchers usually divide China into eight geographic regions: Northeast, North China, East China, Central China, South China, Northwest, Southwest, and Qinghai-Tibet (Fang et al., 2017) for study purpose. Thus, we construct nine data sets, i.e., one national data set and eight regional data sets, by the TBIs geographic locations as per prior research (Fang et al., 2017), to examine, first, which components have significant impacts on the performance of an entrepreneurial ecosystem at the incubator level, and second, are there regional differences among the eight regions in China. In the maps shown in Fig. 2, the circles represent the 33 major cities from our data set, and the eight colors differentiate those cities by the eight culture regions in 2017 (we can provide other years' chart if needed). Size of the circle represents the GDP amount of each city on the left panel and the number of graduated incubatees on the right panel from year 2017. These two charts illustrate the geographical locations of the cities that host our TBIs. In general, we can see three clusters have higher GDPG and higher number graduated incubatees as well, in North China (red circles in Beijing area), East China (orange circles around Shanghai area), and South China (yellow circles around Guangdong area).

Figure 2. GDP and the number of graduated incubatees of the 33 selected cities across eight regions, 2017.

Here, we only present the descriptive statistics of a number of important and interesting variables of the national-level data set and the eight regional data sets, respectively (Table 2). We can provide the full list of all 88 variables and their descriptive statistics based on requirements. Table 2 shows that the average number of graduated incubatees of each incubator in each region varies across the country. The average graduated incubatees of each incubator each year is 6.22 at the national level, and the national median is 4. Among the eight regions, North, Central, and Southwest are higher than the national average, while Northeast, East, South, Northwest, and Qinghai-Tibet are lower. At the national level, the maximum number IP application in a year is 1,320, and it is the highest number an IP application in an incubator could make in a year, from an incubator in the East Region, the most developed region in China. The national average of the number of IP application is 64, and the national median is 30. For the number of IP applications, North, Central, South, Northwest, and Southwest are all higher than the national average. Further, the descriptive statistics vary across regions significantly. Further, the total number of observations of all three years in Northwest, Southwest, and Qinghai-Tibet is 96 (32 incubators), 177 (69 incubators), and 48 (16 incubators), respectively, which are rather small. Thus, we will drop those three regions in our regression analysis because of their limited sample size, which may bias the regression model results.

### 3.2 Methodology

#### 3.2.1 Factor Analysis

Considering the relatively large number of variables (88) in the data set, which would be difficult to interpret or analyze, we apply factor analysis to the entire variable list to decrease the number of variables in order to reduce the complexity of the analysis. Factor analysis has been used by many social science and economic studies to decrease the large number of indicators into a smaller number of dimensions for a straightforward economic interpretation (Fagerberg & Srholec, 2008). The basic idea behind factor analysis is that the variability of many observable variables may reflect a lower number of unobserved factors. We use this insight to build a smaller number of groups of "unobserved factors" for our analysis by constructing the observed variables as linear combinations of the unobserved factors. Therefore, the complexity of the large number of variables can be reduced to lower dimensions.

#### 3.2.2 Fixed Effects vs. Random Effects Panel Regression

Given our uniquely extensive panel data, we can run two types of panel data regression models, i.e., the fixed effects model or the random effects model. The fixed effects model allows us to control for the time-invariant variables that we cannot observe or measure, such the quality of each specific business incubator or an incubator's idiosyncratic heterogeneity. Random effects assume that the entity's error term is not correlated with the explanatory variables, which allows for time-invariant variables to be added to the model. The problem with the random effects model is that some time-invariant variables may not be available, which leads to omitted variable bias in the model. Either a fixed effects model or a random effects model has its own empirical advantages and disadvantages. In practice, a Hausman test can be conducted to distinguish between the random effects and fixed effects models. If the Hausman test is statistically significant, the fixed effects model is preferred over the random effects regression model.

We adopt the number of graduated incubatees from an incubator each year as the dependent variable, which is one of the most common measurements for an incubator's performance in research (Fukugawa, 2018; Harper-Anderson & Lewis, 2018).

## 4.    Results and Discussion

### 4.1 Main Factors

Table 3 presents the results of our factor analysis. The Kaiser-Meyer-Olkin (KMO) test was performed to evaluate the fitness of our factor analysis; the result is 0.775, which means our factor analysis suits this data set well (Kaiser, 1974). Our factor analysis identified seven factors that collectively explained 82.7% of the total variance. Table 3 shows the loadings from our factor analysis, which are the correlation coefficients between the variables (rows) and factors (columns).

The first factor loadings are quite heavy on variables such as the number of staff members of an incubator (0.98), the number of staff members who had higher education (0.98), and the number of incubator staff members who received work skill training (0.36). Those variables are obviously associated with the quantity and quality of a TBI's human capital or the people component. The core services of TBIs are financial consulting, management assistance, and other professional business services, which are obviously dependent on the quantity and quality of the staff of TBIs. Thus, we label this factor as related to the people component, which we the "people-service factor." The second factor shows a strong connection with the number of entrepreneurship advisors (0.51) and the number of contracted professional services (0.41) in an incubator. This factor is still related to the people component, which we call the "people-mentor factor."

The third factor loads heavily on several variables, such as research and development (R&D) investment (0.52), the number of intellectual property applications (0.85), and the number of patents held by all incubatees of an incubator (0.86), which are related to the technology component. Based on our knowledge, the relationship between intellectual property/patents and the performance of an entrepreneurial ecosystem has not been studied in prior research, probably because of the data limitations (Fagerberg & Srholec, 2008). Further, this factor demonstrates a strong relationship with the number of incubatees that received venture capital (0.52) and the number of incubatees listed (0.50), which is consistent with our expectation that an incubatee firm with a higher potency of technology component, i.e., high number of IP applications or patents, is associated with achieving better outcomes, i.e., receiving more venture capital or being publicly listed. In addition, this factor is also lightly related to the external venture capital received by the incubatees (0.38), which can be categorized as the capital component. Thus, we name this the "technology-capital factor" because it is associated with the technology and the capital components.

The fourth factor is heavily loading on the number of graduate firms listed (0.45) on Chinese stock exchange market and the number of incubatees that received venture capital (0.42), which overlap with the third factor and is also associated with another variable, i.e., funds from incubator itself (0.24). Thus, we categorize this factor into the capital component, and we call it the "capital factor" because all the heavy loadings of this factor are related to venture capital or funds or to be listed on the stock exchange market, which indicates a successful capital support as well.

The fifth, the sixth, and the seventh factors are all related to the infrastructure component. The fifth factor loading is more associated with the incubating area (0.44), which belongs to infrastructure component because that is the physical infrastructure of an incubator, and incubators' direct investment from government (0.35), which belongs to a capital component. Thus, the fifth factor is a mixture of two components, the infrastructure and the capital components, which we call the "infrastructure-capital factor."

The sixth factor loading is heavy on GDP per capita (0.65) of the city where the TBIs locate, which is related to a city's local economic infrastructure; thus, we categorize this factor as the infrastructure component of an incubator, which we call the "infrastructure-GDP factor."

The seventh factor exhibits a strong association with the industry diversity index (0.76) and the number of universities in the city (0.74) where the TBI locates. The industry diversity index of a region is related to the local economic infrastructure because, usually, the more diversified a region's industry is, the more mature the region's economic infrastructure is (Quigley,1998). In general, social infrastructure includes hospitals, schools, and universities, which contribute to a region's public life, and the number of colleges and universities is used to indicate how good a region's social infrastructure is in the present study (Pradhan & Rawlings, 2002; Latham & Layton, 2019). Hence, we categorize this factor as the infrastructure component and call it an "infrastructure factor."

### 4.2 Regression Result

Based on the factor analysis result from previous section, we run fixed effects and random effects panel regression models with the seven factors as follows:

$$loggrad_{it} = \alpha + \beta_1 People\_service_{it} + \beta_2 People\_mentor_{it} + \beta_3 Tech\_capital_{it} +$$

$$\beta_4 Capital_{it} + \beta_5 Infra\_capital_{it} + \beta_6 Infra\_GDP_{it} + \beta_7 Infrastructure_{it} + u_{it} \quad (1)$$

where $loggrad_{it}$ is the log of the number of graduated incubatees in incubator $i$ in year $t$ (we take log in order to adjust the abnormal variable values' distribution), $People\_service_{it}$ represents the factor associated with the people-service component of incubator $i$ in year $t$, $People\_mentor_{it}$ represents the factor associated with the people-mentor component of incubator $i$ in year $t$, $Tech\_capital_{it}$ represents the factor associated with the technology-capital component of incubator $i$ in year $t$, $Capital_{it}$ represents the factor associated with the3 capital component of incubator $i$ at year $t$, $Infra\_capital_{it}$ represents factor associated with the mixture of infrastructure and capital component of incubator $i$ at year $t$, $Infra\_GDP_{it}$ represents the factor associated with GDP-related infrastructure component of incubator $i$ at year $t$, $Infrastructure_{it}$ represents the factor associated with the infrastructure component of incubator $i$ in year $t$, and the last term, $u_{it}$, serves as the error term.

Because the descriptive statistics in Table 2 show that some incubators had zero graduated incubatees in certain years, we take the logarithm of the number of graduated incubatees as Eq. (2) to adjust the zeros for the logarithm values:

$$\log grad_{it} = \log (grad_{it} + 1) \quad\quad\quad (2)$$

As previously discussed, the above panel regression model is run with six different data sets, i.e., one data set for the entire country's TBI data and five regional data sets, respectively. The six model results are presented in Table 4. Based on the Hausman tests, we can see that the national model, Northeast, East, and the South models represent the fixed effects models, which means the incubators of those models, the national model, and three regional models had strong and constant fixed effects over the three-year time period. The North and Central regions' models are random effects models, which means the incubators' effects are uncorrelated to individual incubator characteristics. The *R*-squared for all models is satisfactory.

### *4.2.1 National Model*

Figure 3. National regression model result and the hypotheses

Figure 3 summarizes the relationship among the seven factors and the four key components, along with the significance of the seven factors on the seven hypotheses based on the regression model result of the overall data set across China. The model result shows that four out of seven factors related to three components, i.e., people-service factor, people-mentor factor, technology-capital factor, and capital factor, have statistically significant positive effects on the incubator's performance, after controlling for the incubator's fixed effects. This means that, when any of the above four factors of an incubator has a high score, the incubator will have a high number of graduated incubatees. These findings match our expectations based on our authors' practical experiences with the incubators in China as well as extant research from China or other countries.

The regression model result reveals if an incubator has a high people-service factor score, which indicates that an incubator has high-quantity or high-quality staff or if an incubator has a high people-mentor factor score, which means that an incubator has high number of entrepreneurship advisors and high number of contracted professional services; then, the incubator will have more successfully graduated incubatees. This result verifies Hypothesis 1: High-quality/quantity human capital will improve the performance of an entrepreneurial ecosystem.

If an incubator has a high technology-capital factor score, which means an incubator has more IP applications/patents and high investment in R&D, then the incubator will have more successful graduated incubatees. This finding also resonates with our expectations and verifies Hypotheses 2 and 3 on the technology component: The total R&D investment has a positive impact on the performance of the ecosystem, and the number of patents/patent applications of an ecosystem owns has a positive impact on an ecosystem's performance.

If an incubator has a high capital factor score, which means the incubator will have high number of graduate firms listed on the stock market and high number of incubatees that received venture capital from investors, then the incubator will have more graduated incubatees. The positive effects of this capital factor and the previous technology-capital factor are also related to our Hypotheses 4 and 5 on capital component: Capital support from venture capitalists has a positive impact on an ecosystem's performance, and financial support from governments has a positive impact on an ecosystem's performance.

However, the three infrastructure related factors, i.e., infrastructure-capital factor, infrastructure-GDP factor, and infrastructure factor, are not statistically significant in the national model, which may suggest that the basic infrastructure of an incubator, such as the incubator area, or local GDP per capita, or the local industry diversity and the higher education environment, do not have statistically significant impacts on the performance of an entrepreneurial ecosystem or an incubator. It may also mean that China is a large-area country with various levels of infrastructure and economies that dilute potential infrastructure effects on an incubator's performance.

In the present study, the infrastructure-capital factor, indicating the area of an incubator, the government's investment and subsidies in an incubator, and the government' subsides to incubatees, has no statistically significant impacts on an incubator's performance. This makes sense because most technical incubatees nowadays do not need a large office or factory area in order to build a successful business in information technology or biology science; hence, a large incubator area may not play a role in the success of assisting incubatees. However, the interesting finding is that a government's investment in incubators and a government's subsidies to incubatees do not have statistical significant impacts on an incubator's performance, which may be because the significant technology-capital factor also includes a government's investment and subsidies; hence, the impact is probably split between the two factors: technology-capital factor and infrastructure-capital factor. This finding, i.e., that a government's investment or subsidies do not play a role in the success of the TBIs, is consistent with Yin (2009), which revealed that a beneficial government policy was not effective for TBI development in China.

The other two insignificant infrastructure-related factors, i.e., infrastructure-GDP and infrastructure-diversity-education factor, suggest that a more diversified industry or more universities adjacent to a city has no statistically significant impact on the performance of TBIs in this city. The infrastructure-GDP factor, which scores highly on GDP per capita in a city's urban area, has a strong correlation with regional economic development, but it may not tie to a city's innovation driver. One explanation for this phenomenon, according to Marshall (1890), is that regions with a specialized industry, contrary to a diversified industry, tend to be more innovative because knowledge spillover is easier between similar nearby companies. Further, over the last few decades, TBIs have focused on technologies such as the Internet, high tech, biomedicine, etc. Thus, unrelated industry diversity or general traditional industry diversity does not benefit innovation of TBIs (Boschma & Iammarino, 2010). Another explanation may be that China's regional variation is too strong, which effectively dilutes industry diversity in this national data set. Regarding the relationship between the number of universities and innovation within a region, prior research has found that the geographic connection between educational institutions and innovation may no longer hold due to the increased mobility of educated people within a country (Florida, 2002; Florida et al., 2008). Similarly, Xiao and North (2018) found that the number of universities in a city does not have a significant impact on the scale or the type of innovation activity in incubators in China.

*4.2.2 Regional Models*

Figure 4. Five regional regression model results and the hypotheses.

Figure 4 summarizes the regression model results of the five regional data sets from Table 4 and the relationships among the seven factors and the four key components that we developed, along with the relationship among the seven factors and seven hypotheses that we constructed.

The people-service factor has a statistically significant effect in Northeast, East, and Central China but not in North and South China. The people-service factor is heavily loading on the quantity and quality of the incubator staff who provide supportive service work to the incubator's residential startup firms, which has a significant impact on TBI performance and matches our expectation that higher quantity of an incubator' service should lead to more graduated incubatees or a better-performing incubator. This statistical significance is also consistent with our national model result, as the previous session discussed. Regarding why the people-service factor does not have a statistically significant effect in North and South China, one explanation might be that the service market in South China is more developed, and that most of the general services needed by incubatee firms are outsourced, without depending on the internal services provided by the incubators. This assumption will need further study, with more data or more time periods. However, that is beyond the research range of the present study, and it can be one of the future studies. The people-mentor factor shows a statistically significant impact on the incubator performance in two regions, i.e., North and Central China, which suggests that mentor services are greatly needed in those two regions, which are less-developed regions compared with the South, Northeast, and East China regions. This result is also consistent with our national model. Moreover, the people-service factor and the people-mentor factor are related to the people component and Hypothesis 1.

The technology-capital factor is statistically significant and positive in all five regional models; it is the only factor that is statistically significant and positive in all five regional models as well as in the national model. The technology-capital factor is related to the technology component and the capital component; further, it is the only factor related to the technology component. It is exciting to see this factor so consistently significant across all models because the nature of this component is cohesive to the technology and the capital components, which are the keys for success of any technical business or technical business incubator. The significant association between the technology-capital factor and the success of a TBI is a new finding, according to the authors' knowledge, and, to our best knowledge, it has not been discussed in extant studies. This indeed makes sense because the major goal of TBIs is to incubate high-tech startup firms and to help those nascent companies gain maturity or graduate. Of course, capital is one of the major drivers for any business success. The higher the technology-capital factor score of an incubator is indicates that the incubatees of an incubator will own more IP applications and patents as well as invest more in R&D; further, the incubator will have more graduated firms listed on the stock exchange market, more incubatees received venture capital, and the incubatees will receive more venture capital; in addition, the better the performance of the incubator, the incubator will have more graduated incubatee firms. The technology-capital factor's significant positive effects support our Hypotheses 2, 3, and 4.

Another capital component-related factor is the capital factor, which is related to the number of graduated incubatees listed on the stock exchange market and the number of incubatees receiving venture capital. The capital factor exhibits positively statistically significant influence on the performance of an incubator in the Northeast, North, East, and South China, four out of five regions as well as the national model, which means, in an incubator, the more graduated incubatees listed on the stock market and the more incubatees received venture capital, the better performance of an incubator. This finding supports our Hypotheses 4 and 5.

All three factors related to infrastructure component are not statistically significant in the national model, but some are statistically significant in some regional models.

Infrastructure-capital factor shows a statistically significant impact on the performance of an incubator in the East Region only, which is related the capital component and the infrastructure component and supports our Hypotheses 5 and 6. This infrastructure-capital factor is not statistically significant in the national model, however, probably because the uneven development and the diversified infrastructure across China diluted the infrastructure-related effects, which are dependent on the regional economy.

The infrastructure-GDP related factor has a statistically significant impact on the incubator's performance in the Northeast and North regions, which indicates that higher GDP may have an influence on the performance of the incubator's performance, and this finding supports our Hypothesis 6. Again, this infrastructure-GDP factor does not have a statistically significant impact on the performance of the TBIs in the national model.

The infrastructure-diversity-education factor has no statistically significant impact on a TBI's performance in all regions, except for South China, where it has a negative effect. This interesting negative impact may suggest that the cities with more universities or more diversified industry may be less innovative. This factor has no statistically significant impact on TBI's performance in the national model as well as in the other four regional models, which is also consistent with previous research (Boschma & Iammarino, 2010; Xiao & North, 2018). This finding relates to our Hypotheses 6 and 7.

## 5.    Contributions and Limitations

### 5.1 Contributions

The present study makes a few contributions. First, we develop a new framework with four key components to describe an entrepreneurial ecosystem: people, technology, capital, and infrastructure components. To the best of our knowledge, there are no prior theoretical or empirical studies examining the framework of entrepreneurial ecosystems that combine those four components. Previous studies have focused on various forms of the essential elements or components of an entrepreneurial ecosystem, which we call people, capital, or infrastructure (Isenberg, 2010; Feld, 2012; Spigel, 2017; Chen et al., 2019; World Economic Forum, 2013); based on the authors' knowledge, however, research on entrepreneurial ecosystem strains have ignored the technology component. We particularly believe that technology should be an important element to be evaluated in nowadays' entrepreneurship or entrepreneurial ecosystems.

Second, empirically, we find encouraging and exciting results by utilizing a three-year longitudinal data set from MoST, China, which verifies that our proposed framework can be used to examine the relationship among the four key components of an entrepreneurial ecosystem and the success of the entrepreneurial ecosystem. We model a technology business incubator as a mini entrepreneurial ecosystem and investigate how various factors related to the four key components affect the performance of the TBI. We examined the relationships between those factors and the TBI's performance by running our panel regression models on a national data set, which includes all the TBI's data across China as well as on five regional sub-data sets, each of which includes TBIs of a specific region of China, considering China's large geographic area, long uneven economic development history, and diversified local culture.

Our national model shows that all the four factors related to the people component, technology component, and capital component have a statistically significant positive impact on the entrepreneurial ecosystem's performance, while none of the three factors related to infrastructure component have any statistically significant impact on the ecosystem's performance. Among the five regional models, the technology-capital factor related to technology and capital components has a statistically significant positive impact on the ecosystem's performance in all regions. Other factors related to the people component, technology component, capital component, and infrastructure component have varied positive impacts in different regions, while the infrastructure-diversity-education factor related to the infrastructure component has a statistically significant negative impact on the ecosystem's performance in South China, which may need further study with more data.

At last, our research also expands the empirical study on innovations in China, which is the second-largest economy of the world and the largest transforming developing economy in the world. China was a purely central planning economy before 1980s; today, the state-owned economy still plays a large role in the entire economy. In China, government support is typically regarded as one of the most powerful forces controlling the development of the economy because the Chinese government usually has a strong influence over industry due to the dominant state-owned economy. However, interestingly, based on our factor analysis, the government support does not appear to be a statistically significant factor among all the variables. This may suggest that, following about 40 years of economic reform, a market economy may become the primary driver in the development of innovation in China rather than the state-owned economy. This finding is consistent with the findings of Yin (2009), indicating that beneficial government policies were not effective in TBI development in China, and Yin's (2009) study was based on one-time survey data of 136 TBIs in one Chinese province, while the present study is based on a national panel data set collected over a course of three years.

*5.2 Limitations and Future Research*

      This study has limitations. First, the present study is based on empirical data from TBIs in large cities in China, as there are not sufficient data from smaller cities; thus, interpretation of the model results would be limited. The study could further be extended to smaller cities or rural areas if more data become available. Second, another direction for future research would be to conduct onsite interviews to further investigate the relationship between significant factors and an ecosystem's performance and more detailed interpretations beyond the quantitative data. Third, the present study is based on empirical data from China. Could the empirical results be generalizable to other countries or regions that call for further investigation? Yes, the research method can be extended to other countries or other entrepreneurial ecosystems if a panel data set with a similar detailed level data is available; further, how to improve the performance of an entrepreneurial ecosystem to promote technology innovation is a worldwide question pertaining to economic development.

## 6.    Conclusion

The present study proposes an entrepreneurial ecosystem framework with four key components, i.e., people, technology, capital, and infrastructure, then applies the ecosystem framework to examination the performance of technology business incubators in a developing economy. We empirically verify the framework by utilizing a longitudinal survey data set on all the national technology business incubators in China from 2015 to 2017, which includes dozens of variables related to the four key components of all the incubators in major cities. Our factor analysis and panel regression model results show that an entrepreneurial ecosystem's people component, technology component, and capital component have statistically significant impacts on an ecosystem's performance when applied to the overall national data set; further, all four components have various statistically significant impacts on the five regional sub-data sets.

## Declaration of Interest Statement

No potential conflict of interest was reported by the authors.

**Tables**

Table 1.  Descriptive statistics of key variables of our dataset

| | Variables | Number of IP applications by incubatees | Number patents by incubatees | Number of incubator's full time staff | Number of incubator's full time staff who have higher education | Industries diversity index of the city | Number of colleges and universities in the city | Number of graduate firms listed on the stock market | Number of incubatees received venture capital | Number of graduate incubatees every year (dependent variable) |
|---|---|---|---|---|---|---|---|---|---|---|
| All Regions | Observations | 2571 | 2571 | 2571 | 2571 | 2571 | 2571 | 2571 | 2571 | 2571 |
| | Mean | 63.88 | 105.86 | 17.68 | 16.67 | 2.39 | 34.65 | 1.45 | 17.56 | 6.22 |
| | Median | 30 | 46 | 14 | 13 | 2.41 | 31 | 0 | 6 | 4 |
| | Std. Dev | 99.81 | 168.47 | 14.53 | 13.58 | 0.18 | 17.83 | 3.45 | 36.34 | 8.05 |
| | Min | 0 | 0 | 0 | 0 | 1.77 | 6 | 0 | 0 | 0 |
| | Max | 1320 | 2171 | 204 | 198 | 2.71 | 72 | 36 | 542 | 81 |
| Northeast | Observations | 255 | 255 | 255 | 255 | 255 | 255 | 255 | 255 | 255 |
| | Mean | 32.62 | 64.51 | 18.04 | 16.85 | 2.50 | 25.96 | 0.40 | 7.84 | 4.73 |
| | Median | 9 | 17 | 13 | 12 | 2.57 | 27 | 0 | 2 | 3 |
| | Std. Dev | 84.6 | 144.11 | 16.04 | 15.11 | 0.13 | 2.96 | 1.33 | 13.18 | 7.10 |

|  |  |  |  |  |  |  |  |  |  |
|---|---|---|---|---|---|---|---|---|---|
|  | Min | 0 | 0 | 0 | 0 | 2.28 | 20 | 0 | 0 | 0 |
|  | Max | 940 | 1338 | 94 | 94 | 2.65 | 29 | 12 | 61 | 68 |
| North | Observations | 390 | 390 | 390 | 390 | 390 | 390 | 390 | 390 | 390 |
|  | Mean | 74.9 | 130.87 | 20.81 | 19.38 | 2.55 | 46.98 | 2.06 | 23.25 | 8.16 |
|  | Median | 40.5 | 64 | 16 | 15 | 2.7 | 67 | 0 | 10 | 5 |
|  | Std. Dev | 101.4 | 178.89 | 14.83 | 13.73 | 0.18 | 20.84 | 3.88 | 42.61 | 9.35 |
|  | Min | 0 | 0 | 2 | 1 | 2.19 | 10 | 0 | 0 | 0 |
|  | Max | 835 | 1270 | 120 | 120 | 2.71 | 67 | 31 | 453 | 55 |
| East | Observations | 978 | 978 | 978 | 978 | 978 | 978 | 978 | 978 | 978 |
|  | Mean | 57.38 | 90.38 | 15.51 | 14.84 | 2.34 | 31.97 | 0.95 | 15.98 | 5.01 |
|  | Median | 27 | 40 | 12 | 12 | 2.35 | 35 | 0 | 6 | 2 |
|  | Std. Dev | 91.24 | 143.8 | 12.28 | 11.52 | 0.1 | 8.35 | 2.15 | 26.2 | 7.08 |
|  | Min | 0 | 0 | 0 | 0 | 2.00 | 8 | 0 | 0 | 0 |
|  | Max | 1320 | 1873 | 204 | 198 | 2.43 | 38 | 23 | 201 | 81 |
| Central | Observations | 213 | 213 | 213 | 213 | 213 | 213 | 213 | 213 | 213 |
|  | Mean | 91.65 | 135.87 | 21.23 | 20.16 | 2.32 | 32.92 | 3.11 | 27.15 | 9.60 |
|  | Median | 56 | 77 | 17 | 15 | 2.31 | 25 | 1 | 15 | 7 |
|  | Std. Dev | 116.98 | 162.65 | 14.38 | 14.27 | 0.10 | 10.61 | 4.98 | 39.61 | 9.69 |
| | Min | 0 | 0 | 4 | 4 | 2.04 | 23 | 0 | 0 | 0 |
| | Max | 802 | 1105 | 83 | 81 | 2.50 | 46 | 35 | 243 | 61 |
| South | Observations | 417 | 417 | 417 | 417 | 417 | 417 | 417 | 417 | 417 |
| | Mean | 65.66 | 116.8 | 16.00 | 14.68 | 2.27 | 42.42 | 1.55 | 12.85 | 5.53 |
| | Median | 25 | 42 | 12 | 11 | 2.46 | 72 | 0 | 2 | 7 |
| | Std. Dev | 112.07 | 205.65 | 17.97 | 16.12 | 0.27 | 30.4 | 4.09 | 33.66 | 8.35 |
| | Min | 0 | 0 | 0 | 0 | 1.77 | 7 | 0 | 0 | 0 |
| | Max | 1258 | 2171 | 202 | 180 | 2.53 | 72 | 30 | 305 | 61 |
| Northwest | Observations | 96 | 96 | 96 | 96 | 96 | 96 | 96 | 96 | 96 |
| | Mean | 91.4 | 173.14 | 22.18 | 21.34 | 2.51 | 32.63 | 2.76 | 33.45 | 5.69 |
| | Median | 59 | 84 | 20 | 18 | 2.52 | 42 | 1 | 13 | 4 |
| | Std. Dev | 110.23 | 247.29 | 17.84 | 17.66 | 0.05 | 15.15 | 6.2 | 75.56 | 5.89 |
| | Min | 0 | 0 | 0 | 5 | 2.45 | 6 | 0 | 0 | 0 |
| | Max | 480 | 1306 | 120 | 120 | 2.63 | 42 | 36 | 430 | 32 |
| Southwest | Observations | 177 | 177 | 177 | 177 | 177 | 177 | 177 | 177 | 177 |
| | Mean | 77.33 | 113.68 | 19.14 | 18.29 | 2.42 | 24.10 | 1.62 | 21.02 | 8.89 |
| | Median | 44 | 59 | 17 | 26 | 2.46 | 25 | 0 | 6 | 6 |
| | Std. Dev | 95.72 | 150.61 | 10.47 | 9.8 | 0.09 | 3.15 | 3.47 | 54.32 | 9.40 |

|  |  |  |  |  |  |  |  |  |  |
|---|---|---|---|---|---|---|---|---|---|
|  | Min | 0 | 0 | 2 | 2 | 2.22 | 18 | 0 | 0 | 0 |
|  | Max | 571 | 776 | 59 | 55 | 2.55 | 27 | 24 | 542 | 51 |
| Qinghai-Tibet | Observations | 48 | 48 | 48 | 48 | 48 | 48 | 48 | 48 | 48 |
|  | Mean | 27.07 | 42.49 | 18.69 | 17.4 | 2.50 | 17 | 0.96 | 8.42 | 5 |
|  | Median | 7 | 9 | 20 | 18 | 2.5 | 17 | 0 | 4 | 3 |
|  | Std. Dev | 45.2 | 71.83 | 7.21 | 7.32 | 0.01 | 0 | 2.65 | 14.32 | 7.54 |
|  | Min | 0 | 0 | 2 | 2 | 2.49 | 17 | 0 | 0 | 0 |
|  | Max | 175 | 298 | 34 | 30 | 2.50 | 17 | 11 | 60 | 39 |

Table 2. Result of factor analysis

| Factors | People-service | People-mentor | Technology-capital | Capital | Infrastructure-capital | Infrastructure-GDP | Infrastructure-diversity-education |
|---|---|---|---|---|---|---|---|
| Incubator's investment on shared technology platform | 0.15 | -0.02 | 0.02 | 0.07 | 0.11 | 0.01 | -0.04 |
| Incubatees' investment on R&D | 0.13 | 0.05 | **0.52** | 0.10 | 0.15 | 0.08 | -0.01 |
| Number of IP applications by incubatees | 0.15 | 0.03 | **0.85** | 0.01 | -0.001 | -0.01 | -0.03 |
| Number of patents by incubatees | 0.16 | 0.05 | **0.86** | 0.04 | 0.05 | 0.002 | -0.03 |
| Number of purchased abroad patents | 0.03 | 0.02 | 0.19 | 0.09 | 0.02 | 0.06 | -0.04 |
| Number of national level R&D projects | 0.01 | -0.03 | 0.03 | 0.02 | 0.04 | -0.04 | 0.001 |
| Number of incubator's full time staff | **0.98** | 0.01 | 0.11 | 0.03 | 0.03 | 0.01 | 0.01 |
| Number of incubator's full time staff who havehigher education | **0.98** | 0.04 | 0.14 | 0.03 | 0.01 | -0.02 | 0.001 |
| Number of incubator's staff who received skill training | **0.36** | 0.16 | 0.26 | -0.01 | -0.04 | -0.06 | 0.01 |
| Average graduate periods of incubatees | 0.10 | 0.08 | 0.29 | 0.12 | 0.02 | -0.08 | -0.02 |
| Number of incubatees | 0.13 | 0.22 | 0.31 | 0.11 | 0.21 | -0.03 | 0.02 |
| Incubated fund from incubator | 0.16 | 0.10 | 0.16 | 0.24 | 0.05 | 0.04 | 0.04 |
| External venture capital received by the incubatees | 0.11 | 0.06 | **0.38** | 0.14 | -0.07 | 0.06 | 0.13 |
| Incubating area | 0.15 | 0.03 | 0.34 | 0.12 | **0.44** | -0.07 | -0.06 |
| Incubators'direct investment from government | 0.10 | 0.07 | 0.24 | 0.11 | **0.35** | -0.06 | -0.06 |
| Incubators' subsidies from government | 0.21 | 0.05 | 0.20 | 0.14 | 0.14 | 0.10 | -0.02 |
| Incubators' tax reduction from government | 0.14 | 0.17 | 0.15 | 0.05 | 0.001 | -0.0001 | -0.06 |
| Incubatees' subsidies from government | 0.08 | 0.08 | 0.34 | 0.20 | 0.13 | 0.06 | -0.09 |
| Number of entrepreneurship advisors | 0.13 | **0.51** | 0.21 | 0.08 | -0.01 | -0.02 | 0.01 |
| Number of contracted professional services | 0.13 | **0.41** | 0.29 | 0.11 | 0.13 | -0.04 | 0.03 |
| Number of training sessions for incubatees | 0.10 | 0.24 | 0.10 | 0.12 | -0.02 | 0.01 | -0.001 |
| Number of graduate firms listed on the stock mareket | 0.25 | 0.06 | **0.50** | **0.45** | 0.15 | 0.01 | 0.04 |
| Number of incubatees which received venture capital | 0.15 | 0.23 | **0.52** | **0.42** | 0.04 | 0.01 | 0.02 |
| GDP per capita of the city's urban area | -0.02 | -0.01 | 0.01 | 0.01 | -0.03 | **0.65** | -0.15 |
| Industry diversity index of the city | 0.04 | 0.03 | -0.06 | -0.01 | 0.003 | -0.34 | **0.76** |
| Number of colleges and universities of the city | -0.02 | -0.002 | 0.003 | 0.02 | -0.03 | 0.20 | **0.74** |

Note: The number of observations = 2571, pooled data for 857 incubators over three years.

Table 3.  Results of Panel Regression Models

| | People-service | People-mentor | Technology-capital | Capital | Infrastructure-capital | Infrastructure-GDP | Infrastructure-diversity-education | Constant | Hausman Test | FE/RE | Over-all $R^2$ | # of ob. / incubators | # of cities |
|---|---|---|---|---|---|---|---|---|---|---|---|---|---|
| **All Regions** | 0.28** (0.11) | 0.40** (0.14) | 0.62** (0.12) | 0.74** (0.17) | 0.13 (0.17) | -0.15 (0.12) | -0.23 (0.39) | 0.08* (0.04) | 0.005 | Fixed effects | 0.17 | 2571/ 857 | 33 |
| **Northeast** | 1.53** (0.48) | 0.88 (0.71) | 2.31* (0.94) | 3.82** (1.23) | -0.24 (0.72) | -2.71* (1.15) | 0.27 (1.5) | -0.86 (0.95) | 0.030 | Fixed effects | 0.09 | 255/ 85 | 4 |
| **North** | 0.09 (0.14) | 0.57** (0.18) | 0.69** (0.13) | 0.35* (0.19) | 0.2 (0.2) | -0.42* (0.24) | 0.05 (0.16) | 0.74** (0.19) | 0.480 | Random effects | 0.16 | 390/ 130 | 6 |
| **East** | 0.33* (0.16) | 0.45 (0.3) | 0.97** (0.26) | 0.75* (0.38) | 0.97** (0.39) | 0.16 (0.37) | -0.56 (0.94) | -0.47** (0.94) | 0.050 | Fixed effects | 0.26 | 978/ 326 | 6 |
| **Central** | 0.37** (0.14) | 0.49* (0.24) | 0.55** (0.14) | 0.17 (0.18) | -0.02 (0.21) | -0.22 (0.41) | 0.48 (0.38) | 1.37** (0.18) | 0.659 | Random effects | 0.18 | 213/ 71 | 4 |
| **South** | 0.67 (0.4) | 0.48 (0.4) | 1.07** (0.31) | 1.06* (0.44) | 0.31 (0.7) | -0.16 (0.16) | −3.79** (1.38) | -0.54** (0.2) | 0.015 | Fixed effects | 0.14 | 417/ 139 | 5 |

Notes: * indicates $p < 0.05$ and ** indicates $p < 0.01$.

Figure 1. Number of TBIs in China from 1995–2017

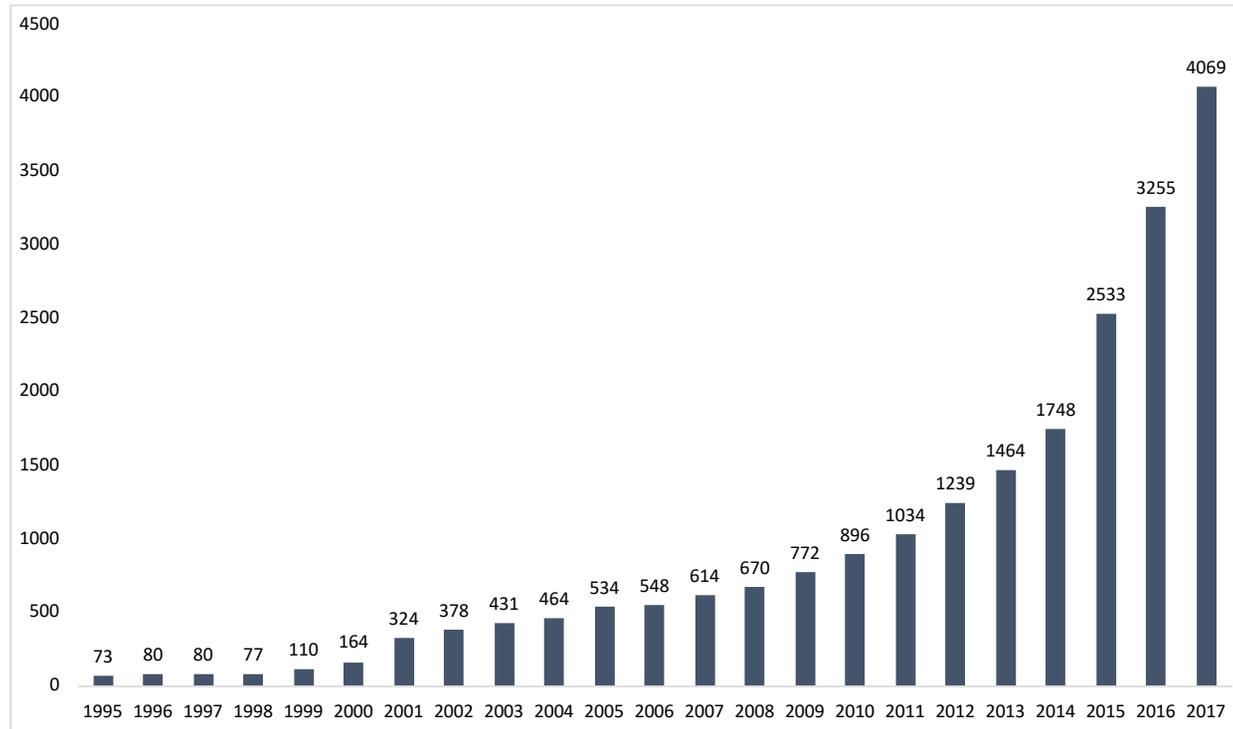

Figure 2. GDP and the number of graduated incubatees of the 33 selected cities across 8 regions, 2017.

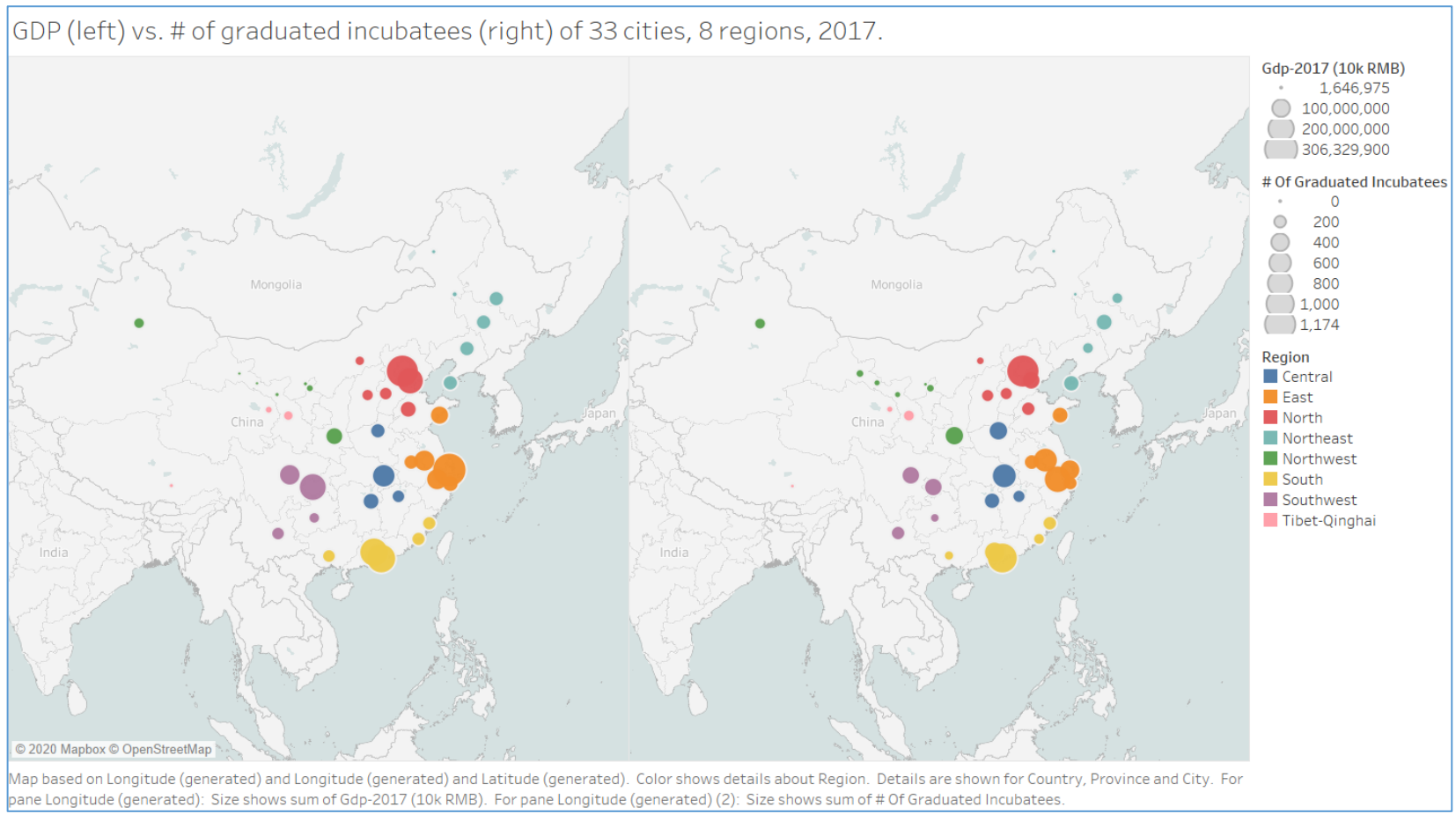

Figure 3. The national regression model result and the hypotheses

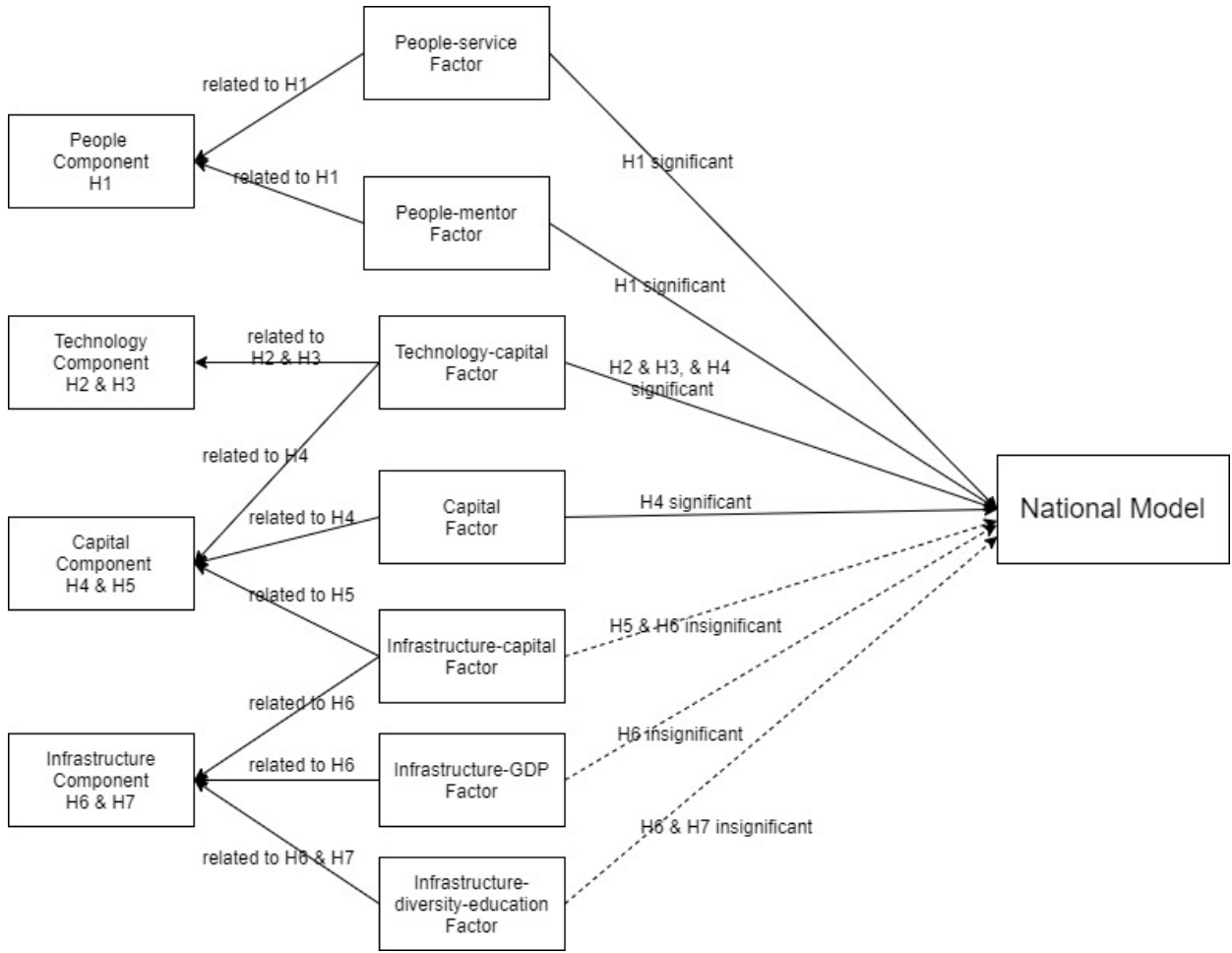

Figure 4. Five regional regression model results and the hypotheses

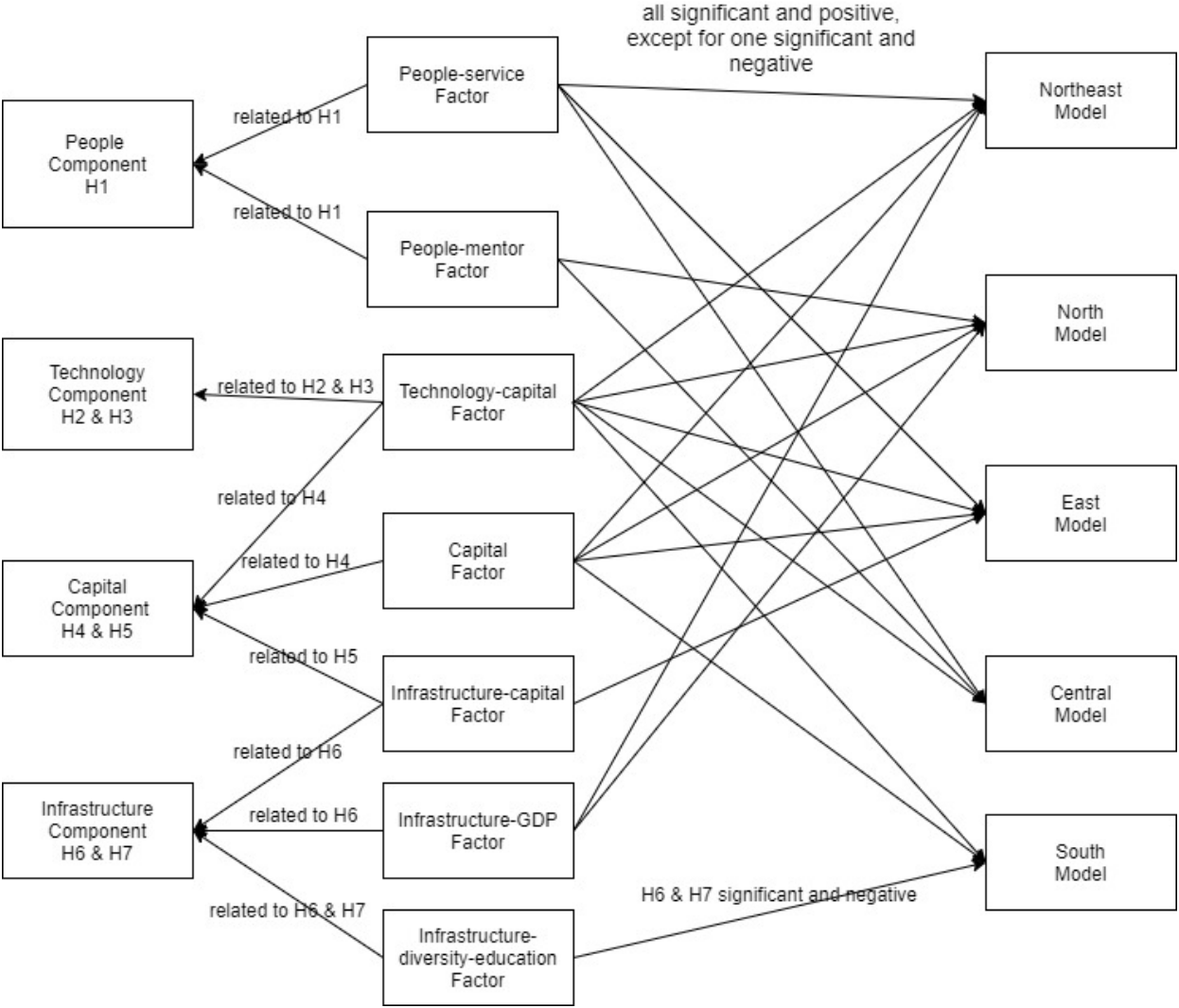

Figure 1. Number of TBIs in China from 1995–2017

Figure 2. GDP and the number of graduated incubatees of the 33 selected cities across 8 regions, 2017.

Figure 3. The national regression model result and the hypotheses

Figure 4. Five regional regression model results and the hypotheses

Word Count: 8,579, excluding tables, figures, and references.